\documentstyle[prd,aps]{revtex}
\input epsf.tex
\def\DESepsf(#1 width #2){\epsfxsize=#2 \epsfbox{#1}}
\begin{document}
\draft
\twocolumn[\hsize\textwidth\columnwidth\hsize\csname
@twocolumnfalse\endcsname

\preprint{\vbox{\hbox{UCSB-HEP-00-05}
\hbox{UMD-PP-01-020}}}

\title{ Large Extra Dimensions, Sterile neutrinos and Solar Neutrino Data}

\author{D. O. Caldwell$^1$\footnote{e-mail:caldwell@slac.stanford.edu},
R. N. Mohapatra$^2$\footnote{e-mail:rmohapat@physics.umd.edu},
and
S. J. Yellin$^{1}$\footnote{e-mail:yellin@slac.stanford.edu} }

\address{$^1$ Department of Physics, University of California, Santa Barbara,
CA, 93106, USA.\\
$^2$ Department of
Physics, University of Maryland, College Park, MD, 20742, USA.}

\date{November 17, 2000}

\maketitle

\begin{abstract}
Solar, atmospheric and LSND neutrino oscillation results require a
light sterile neutrino, $\nu_B$, which can exist in the bulk of
extra dimensions.  Solar $\nu_e$, confined to the brane, can oscillate
in the vacuum to the zero mode of $\nu_B$ and via successive MSW
transitions to Kaluza-Klein states of $\nu_B$.  This
new way to fit solar data is provided by both low and intermediate string
scale models.  From average rates seen in the three types of solar
experiments, the Super-Kamiokande spectrum is predicted with 73\%
probability, but dips characteristic of the $0.06$ mm
extra dimension should be seen in the SNO spectrum.
% PACS: 14.60.Pq; 14.60.St; 11,10.Kk;
\end{abstract}
\vskip2pc]
The four-neutrino scheme in which the solar $\nu_e$ deficit is
explained by $\nu_e\rightarrow\nu_s$ (where $\nu_s$ is a sterile
neutrino), the atmospheric $\nu_\mu/\nu_e$ anomaly is
attributed to $\nu_\mu\rightarrow\nu_\tau$, and the heavier $\nu_\mu$
and $\nu_\tau$ share the role of hot dark matter was originally proposed
\cite{fournu} in order to explain those three phenomena.  Later the
LSND experiment \cite{LSND}, which observed $\bar\nu_\mu\rightarrow
\bar\nu_e$, provided a measure of the mass difference between the nearly
degenerate $\nu_e-\nu_s$ and $\nu_\mu-\nu_\tau$ pairs and required the
three mass differences that were already present in that neutrino
scheme.  Exactly this same pattern of neutrino masses and mixings appears
necessary to allow production of heavy elements
(A $\buildrel >\over \sim$ 100) by type II supernovae\cite{rprocess}.
While qualitatively this neutrino scheme seems to
explain all existing neutrino phenomena, solar neutrino observations are now
sufficiently constraining that the small-angle MSW $\nu_e\rightarrow\nu_s$
explanation appears to be in some difficulty\cite{SK2001}.
Although providing better fits
to the solar data, even active-active transitions in a three-neutrino scheme
do not give a quantitatively good explanation of those data.
In this Letter we point out that there is
a way to achieve an excellent fit and rescue the apparently needed
four-neutrino scheme using large extra dimensions. This is motivated by
the latest developments in string theories, which have made plausible the
interesting possibility that
such large extra space dimensions
($\buildrel < \over \sim$mm)\cite{nima} can exist in conjunction with
low or intermediate fundamental string scales, $M_*$.
Using such models and the average rates from the three types of solar
experiments, we predict the energy spectra from the
Super-Kamiokande\cite{SK2001} and SNO experiments employing
a combination of vacuum and MSW oscillations.  The neutrino-electron
scattering results from Super-Kamiokande is predicted with a $\chi^2$
corresponding to 73\% probability, while the better energy resolution
of the charged-current interactions from SNO should show the predicted dips
in the spectrum.

We have found two classes of models giving the desired phenomenology
(see \cite{cmyprd} for details).  The first model has low $M_*$, as
has been used to give an
alternative to the conventional high-scale SUSY GUT theories in solving the
gauge hierarchy problem and providing low energy signals of string theories.
Since there are no large scales to
implement the conventional seesaw mechanism,
a way to get small neutrino masses is to include singlet
neutrinos in the bulk, in combination with the assumption\cite{dienes} of
an effective global $B-L$ symmetry in the theory below $M_*$.
The large size of the bulk
suppresses neutrino masses to the desired level.
The low Kaluza-Klein (``KK'') excitations have a mass
$\sim R^{-1}\sim 10^{-3} $ eV, which not only provide a natural
way to understand the lightness of the sterile neutrino\cite{apl}, but also
give extra sterile neutrino states in the mass range that can
influence the
shape of the solar neutrino spectrum\cite{dvali}.

This model has
one bulk neutrino, $\nu_B(x,y)$, $y$ being the coordinate of the
fifth dimension, and the standard model on the brand.
% Our considerations apply even if the bulk is six or
%higher dimensional, as long as
%only one extra dimension is large.
%% and all others small
The  $\nu_B(x, y)$ is assumed to couple to the lepton doublet of
the standard model $L$ and of course have a five-dimensional kinetic
energy term.
% The effective 4-dimensional Lagrangian for this system can be
%written as:
%\begin{eqnarray}
% {\cal L} = \kappa  \bar{L} H \nu_{BR}(x, y=0) + \int dy\
% \bar{\nu}_{BL}(x,y)\partial_5 \nu_{BR}(x,y) + h.c.,
% \label{l1}
% \end{eqnarray}
%where from the five-dimensional kinetic energy, we have only kept the
%5th component that contributes to the mass terms of the KK modes in the
%brane; $H$ denotes the Higgs doublet, and 
After electroweak symmetry breaking at scale $v_{wk}$, the $\nu_e-\nu_{B,R}$
coupling leads
to $m_{\nu} = h {M_*v_{wk}\over M_{P\ell}}\sim 10^{-5}$ eV,
where $M_{P\ell}$ is the Planck mass and $h$ is the Yukawa coupling.
Note that this suppression is independent of the number and radius
hierarchy of the extra dimensions, provided that $\nu_B$
propagates in the whole bulk.  Even if the bulk is six or higher
dimensional, there is only one mm-scale dimension.
The smaller dimensions will
contribute only to the relationship between $M_{P\ell}$ and $M_*$,
but their KK excitations will be
very heavy and decouple from the neutrino spectrum, making
the analysis as in five dimensions.

The direct Dirac or Majorana mass terms for the bulk neutrino
can be forbidden by an appropriate choice of geometry and dimension of the
bulk in which the $\nu_B$ resides, making an ultralight $\nu_s$ natural.
For instance, in 5 dimensions the
$Z_2$ orbifold symmetry under which $y\rightarrow -y$ combined with $B-L$
symmetry guarantee this for the conventional definition of
charge conjugation.
 
In order to fit neutrino data, one needs to include new physics in the
brane that will generate a Majorana mass matrix for the three standard
model neutrinos of the form ${ \delta}_{ab}$ (where $a,b= e,\mu,\tau$).
For $\delta_{\mu\tau}$ much bigger than the
other elements, the $\nu_{\mu,\tau}$ in effect decouple from
the $\nu_{e,s}$ and do not affect the
mixing between the bulk neutrino modes and the $\nu_e$. Further, this leads
to maximal mixing in the $\mu-\tau$ sector, as is needed to understand the
atmospheric neutrino data. If $\delta_{\mu\tau}\sim $eV, then
this provides an explanation of the LSND observations. One way to generate
this pattern is to consider an $L_e+L_{\mu}-L_{\tau}$ symmetric extension
of the standard model with additional doubly-charged singlet ($Y=4$) 
and $Y=2$ triplet scalar fields. In the rest of the
Letter, we focus only on the $\nu_e-\nu_s$ sector and how we fit
the solar neutrino data.

The mass matrix for the $\nu_e$ , $\nu_s$
sector can be written
 \begin{eqnarray}
   ({\nu}_{e} \nu_{0B} {\nu}'_{B,-} \nu'_{B,+})\left(\begin{array}{cccc}
\delta_{ee} & m &\sqrt{2} m & 0 \\ m &0 & 0 & 0\\ \sqrt{2} m & 0 
& 0 & \partial_5\\ 0 & 0 & \partial_5 &0
 \end{array}\right)\left(\begin{array}{c}\nu_e \\ \nu_{0B} \\
 \nu'_{B,-} \\ \nu'_{B,+}\end{array}\right),
 \label{m1}
 \end{eqnarray}
where $\nu'_B$ represents the KK excitations, and
the off-diagonal term  $\sqrt{2} m$  is actually an infinite row vector
of the form $\sqrt{2} m (1,1,\cdots)$.  The operator $\partial_5$ stands
for the diagonal and infinite KK mass matrix whose $n$-th entry is given by
$n/R$.  Using this short-hand notation makes it easier to calculate the exact
eigenvalues $m_n$ and the eigenstates of this mass matrix,
\begin{eqnarray}
m_n= \delta_{ee} + \pi m^2R\, {\rm cot}(\pi m_nR). \label{characteristic4}
\end{eqnarray}
%\be
% 2 \lambda_n =2\delta_{ee}R+ \pi \xi^2 \cot(\pi \lambda_n),
% \label{char1}
% \ee
%with $\lambda_n=m_nR$, $\xi=\sqrt{2}mR$, and 
The equation for eigenstates is
% \be 
% \tilde \nu_{nL} = {1\over N_n} \left[ \nu_L + 
% {\sqrt{2} m \partial_5\over m_n^2 - \partial_5^2 }\, \nu'_{BL}\right],
% \label{nus}
% \ee
%From this it follows that
\begin{eqnarray}
N_n\tilde \nu_n = \nu_e + {m\over m_n}\nu_{0B}
+ {\sqrt{2} m 
\left(m_n\nu'_{B,-} + \partial_5 \nu'_{B,+}\right)\over m_n^2 - \partial_5^2 },
\label{nus}
\end{eqnarray}
where the sum over the KK modes in the last term is implicit. 
$N_n$ is the normalization factor given by
\begin{eqnarray}
N_n^2 = 1 + m^2\pi^2R^2 + {(m_n-\delta_{ee})^2\over m^2}.
\label{Nn}
\end{eqnarray}
% \be
%  N^2_n = {2\over\xi^2}\left(\lambda_n^2 + f(\xi)\right),
% \label{Nn}
%  \ee
%where $f(\xi)= \xi^2/2 + \pi^2 \xi^4/4$.
%It will depend on the value of $\xi$. 
%We will
%be interested in the small $\xi$ region of the parameter space. In this
%case it is easy to see that the coupling of the nth mode is given by
%$\xi/n$. 
Note that in the limit of $\delta_{ee}=0$, the $\nu_e$ and $\nu_{B,0}$ are
two, two-component spinors that form a Dirac fermion with mass m.
The KK modes come in pairs of mass $m_n=\pm k_n/R$, with
$k_n$ a positive integer, and they couple to the
$\nu_e$ approximately as $mR$.
Once we
include the effect of $\delta_{ee}\neq 0$, they become Majorana fermions
with masses given by $m_1\approx +\delta_{ee}/2 + m $ and $m_2\approx 
+\delta_{ee}/2 - m$, and they are maximally mixed; i.e., the two mass
eigenstates are $\nu_{1,2}\simeq \frac{\nu_e \pm \nu_{B,0}}{\sqrt{2}}$.
Thus as the $\nu_e$ produced in a weak interaction process evolves, it
oscillates to the $\nu_{B,0}$ state with an oscillation length $D\simeq
E/(2m\delta_{ee})$, which for natural
values of $m,\delta_{ee}$ gives $D$ of order of the Sun-Earth
distance so that our model leads to vacuum oscillation (``VO'') of the solar
neutrinos.
Furthermore, since the $\nu_e$ also mixes with the KK modes of the bulk
neutrinos with a $\delta m^2 \sim 10^{-5}$ eV$^2$, this
brings in the MSW resonance transition of $\nu_e$ to $\nu_{B,KK}$
modes at higher energies.

The second way to achieve the same phenomenology is to use a much
higher string scale associated with local $ SU(2)_L\times
SU(2)_R\times U(1)_{B-L}$ symmetry of the left-right symmetric model in 
 the brane, coupled to a single $\nu_B$ in the large
fifth dimensional bulk.
Using the usual fermion content
of the left-right models and breaking the right-handed symmetry by Higgs
doublets, one obtains the seesaw type matrix in the presence of the
infinite tower of KK modes. The heavy right-handed neutrino now decouples,
leading to the following neutrino mass matrix in the one generation case
(the basis is  $ ({\nu}_{e} \nu_{0B} {\nu}'_{B,-} \nu'_{B,+})$)
 \begin{eqnarray}
\frac{1}{M}\left(\begin{array}{cccc}
m^2+\Delta & m\alpha &\sqrt{2} m\alpha & 0 \\ m\alpha &\alpha^2 &
\sqrt{2}\alpha^2 & 0\\ \sqrt{2} m\alpha & \sqrt{2}\alpha^2
& 0 & \partial_5\\ 0 & 0 & \partial_5 &0
 \end{array}\right),
 \end{eqnarray}
where $\alpha\simeq \frac{ h_{\ell}M_*v_R}{M_{P\ell}}$, $m=h_\ell
v_{wk}$, $M=v^2_R/M_*$, and $v_R$ is the scale of $SU(2)_R$ breaking.
$\Delta$ denotes the radiative correction
induced due to the extrapolation from the $v_R$ scale to the weak scale. 
The lowest eigenvalue of this mass matrix is $\sim
\frac{\alpha^2\Delta}{M(\alpha^2+m^2)}$, and the 
 next lowest one is $\frac{m^2+\alpha^2}{M}$. For
$m,\alpha \sim 1-5 $ MeV (similar to the first generation fermion
mass) and $M\simeq 10^{9}$ GeV, we get this eigenvalue to be of order
$10^{-6}-2.5 \times 10^{-5}$ eV. Its square is therefore in the range
where the VO solution to the solar neutrino puzzle can be applied. Also
for $m\simeq \alpha$, the mixing angle between the zero eigenvalue mode
and this mode is maximal. Thus this model has properties similar to the 
first model for neutrinos, and below we carry out our fit to solar neutrino
data using the latter.

For propagation in solar matter, the eigenvectors and eigenvalues can be
found by replacing the squared mass matrix, $M^2$, for the neutrinos,
with $M^2 + H$, where $H=2E\rho_e$ when acting on $\nu_e$ (where
$\rho_e=\sqrt{2}G_F(n_e-n_n/2)$),
and zero on sterile neutrinos.  Defining
\begin{eqnarray}
w={E\rho_e\over m_n\delta_{ee}} + \sqrt{1 + 
\left({E\rho_e\over m_n\delta_{ee}}\right)^2},
\label{wformula}
\end{eqnarray}
($w=1$ in vacuum) Eq. \ref{characteristic4} becomes
\begin{eqnarray}
m_n= w\delta_{ee} + \pi m^2R\, {\rm cot}(\pi m_nR). \label{characteristic5}
\end{eqnarray}
The eigenvectors are as in Eq. \ref{nus}, except the
% with the
%difference that the
% coefficients of $\nu_{0B}$ and $\nu'_{B,-}$ 
third term acquires an additional factor $1/w$.
% and the normalization becomes 
%$N_n^2 = 1 + {(1+{1\over w^2})\over 2}\left(\pi^2m^2R^2 +
%{(m_n-w\delta_{ee})^2\over m^2}\right) - {(1-{1\over w^2})\over 2}
%{(m_n-w\delta_{ee})\over m}$.

\begin{figure}[htbp]
\epsfxsize=3in
\begin{center}
\leavevmode
\epsfbox{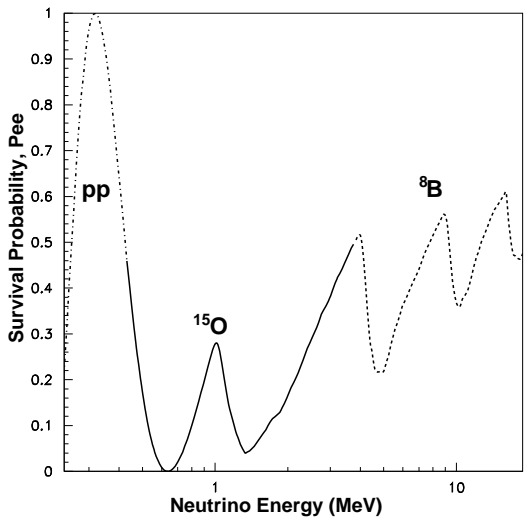}
\end{center}
\caption{Energy dependence of the $\nu_e$ survival probability when
$R=58\,\mu$m, $mR=0.0094$, and
$\delta_{ee}=0.84\times 10^{-7}$ eV.  The dot-dashed part of the curve
assumes the radial dependence in
the Sun for neutrinos from the pp reaction, the solid part assumes $^{15}O$
radial dependence, and the dashed part assumes $^8B$ radial dependence.}
\label{fig:edep}
\end{figure}

In fitting the solar data using VO,
the strategy generally employed is to suppress $^7$Be neutrinos while
also reducing the $^8$B neutrinos by
more than 50\%. The water data then requires an additional contribution, which,
in the case of active VO, is provided by the
neutral current cross section amounting to about 16\% of the charged
current one. Thus in a pure two-neutrino oscillation
picture, VO does not work for active to sterile oscillation.
In our model, however, both vacuum oscillations and MSW oscillations are
important,  since the lowest mass pair of neutrinos is split by a very
small mass difference, whereas the KK states have to be separated by
$>10^{-3}$ eV because of the limits from gravity experiments.
%when $\mu_0\approx 2\times 10^{-3}$ eV, $m_0 \approx 0.05\mu_0
%\approx 10^{-4}$ eV, and $\delta_{ee}\approx (3/m_0)\times 10^{-11} \approx
%3\times 10^{-7}$ eV. 
We can use the first node of the survival probability, $P_{ee}$, to
suppress the $^7$Be.  Going up in energy toward $^8$B neutrinos, $P_{ee}$,
which in the VO case would have
risen to very near one, is reduced by the
small-angle MSW transitions to the different KK excitations of the bulk
sterile neutrinos, as is clear from Fig. \ref{fig:edep}.
%Most dips in Fig. \ref{fig:edep} are a consequence of the MSW resonances of
%successive KK states.
This is a new way to fit the solar neutrino data in models with
large extra dimensions.

To do the fit, we studied the time evolution of the $\nu_e$ state
with a program supplied by W. Haxton\cite{Haxton}, but updated
to use a recent solar model\cite{BP98} and modified
to do all neutrino transport within the
Sun numerically, using no adiabatic approximation.
Changes were also necessary for oscillations into sterile neutrinos and
to generalize beyond the two-neutrino model, for up to
14 neutrinos contribute for the solutions we considered.

For comparison with experimental results, tables of detector sensitivity
for the Chlorine and Gallium experiments were taken from Bahcall's web
site\cite{BP98}.
The Super-Kamiokande detector sensitivity was modeled
using \cite{Nakahata} for the 
resolution in the signal from Cherenkov light.
More details are given elsewhere\cite{cmyprd}.  
Calculations of $P_{ee}$, averaged over the
response of detectors, were compared with measurements.  While theoretical
uncertainties in the solar model and detector response were included 
in the computation of $\chi^2$ as
described in Ref. \cite{Fogli}, the measurement results given here include only
experimental statistical and systematic errors added in quadrature.
The Chlorine $P_{ee}$ from Homestake\cite{Homestake} is
$0.332\pm 0.030$.
Gallium results\cite{N2000} for SAGE, GALLEX and GNO were combined to give a
$P_{ee}$ of $0.579\pm 0.039$.
The $5.0-20$ MeV, 1258-day Super-K experimental $P_{ee}$\cite{SK2001}
is $0.451\pm 0.016$.
The best fits were with $R\approx 58\,\mu m$, $mR$ around $0.0094$, and
$\delta_{ee}\sim 0.84\times 10^{-7}$ eV, corresponding to
$\delta m^2\sim 0.53\times 10^{-11}$ eV$^2$, giving
average $P_{ee}$ for Chlorine, Gallium,
and water of %0.376, 0.512, and 0.463,  0.382, 0.533, 0.454,
%0.386, 0.533, and 0.460,
0.383, 0.533, and 0.450,
respectively, and the $P_{ee}$
energy dependence shown in Fig. \ref{fig:edep}.
For two-neutrino oscillations, the mixing angle is
% coupling between $\nu_e$
%and the higher mass neutrino eigenstate is given by
$sin^22\theta$, whereas here the coupling between $\nu_e$
and the first KK excitation replaces $sin^22\theta$ by $4m^2R^2 = 0.00035$.

Vacuum oscillations between the lowest two mass eigenstates nearly eliminate
electron neutrinos with energies of 
0.63 MeV/(2n+1) for n = 0, 1, 2, ... .  Thus Fig. \ref{fig:edep}
shows nearly zero $P_{ee}$
near 0.63 MeV, partly eliminating the $^7Be$
contribution at $0.862$ MeV, and giving a dip at the lowest neutrino
energy.
%COMMENT: The next 5 lines should be in the PRD, but perhaps should be dropped here.
%Increasing $\delta_{ee}$ moves the low energy dip to the right
%into Gallium's most sensitive pp energy range, making the fit worse.
%Decreasing $\delta_{ee}$ increases Gallium,
%but hurts the Chlorine fit by moving the higher energy
%vacuum oscillation dip further to the left of the $^7Be$ peak.
MSW resonances with mass pairs of higher KK states start causing the
third and fourth eigenstates to be
significantly occupied above $\sim 0.8$ MeV, the fifth and sixth eigenstates
above $\sim 3.7$ MeV, the 7'th and 8'th above $\sim 8.6$ MeV, and the 9'th
and 10'th above $\sim 15.2$ MeV.
Fig. \ref{fig:edep} shows dips in $P_{ee}$
just above these energy thresholds.
%COMMENT: Another example of something that should be in PRD but is a place where
%the PRL can be made shorter:
%The typical values of the survival probability within the
%$^8B$ region ($\sim 6$ to $\sim 14$ MeV) are quite sensitive to the value of
%$mR$.  As can be seen from Eq. \ref{Nn}, higher $mR$
%increases $1/N\approx m/m_n\approx mR/n$ for various
%$n$, and thereby increases $\nu_e$ coupling to higher mass eigenstates,
%strengthens MSW resonances, and lowers $\nu_e$ survival probability. 

\begin{figure}[htbp]
\epsfxsize=3in
\begin{center}
\leavevmode
%\epsffile{skspectrum.eps}
\epsfbox{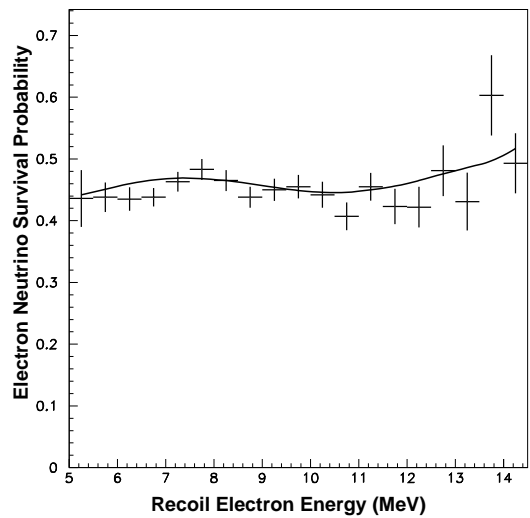}
\end{center}
%\caption{Super-Kamiokande energy spectrum: measured\protect\cite{SK2001}
%results based on 1258 days (error bars)
%and predicted (curve) for the same parameters as in Fig. \ref{fig:edep}.
%The curve is not a fit to these data.}
\caption{Super-Kamiokande 1258-day measured\protect\cite{SK2001}
energy spectrum (error bars)
and predicted (curve) using the parameters of Fig. \ref{fig:edep}
without fitting these data.}
\label{fig:skspect}
\end{figure}

The expected energy dependence of $P_{ee}$ is
compared with Super-K data\cite{SK2001}
in Fig. \ref{fig:skspect}.  The uncertainties are statistical only.
The parameters used in making Fig. \ref{fig:skspect} were chosen to provide
a good fit to the total rates only; they were not adjusted to fit
this spectrum.  Combining spectrum data with rates using the method
described in Ref. \cite{GHPV} gives $\chi^2=14.0$
for the spectrum predicted from the fit to total rates.  With 18 degrees of
freedom, the probability of $\chi^2>14.0$ is 73\%.  A fit
with $\delta_{ee}$ constrained to be very small to
eliminate vacuum oscillations increased the best fit
$\chi^2$ from 3.4 to 4.4.  The same parameters then used
with the Super-K spectrum gave $\chi^2=18.7$ (probability 41\%).  This is
comparable to $\chi^2=19.0$ for an energy independent spectrum.

The seasonal effect was computed for a few points on the Earth's orbit.
If $r$ is the distance between the Earth and the Sun,
${r_0\over r} = 1 + \epsilon\,cos(\theta-\theta_0)$,
where $r_0$ is one astronomical unit, $\epsilon=0.0167$ is the orbital
eccentricity, and $\theta-\theta_0\approx 2\pi(t-t_0)$, with $t$ in years and
$t_0\ =$ January 2, 4h 52m.  Table I
shows very small seasonal variation.

Not only is the seasonal effect very difficult to observe, but also the
smallness of the mixing angle makes day-night effects hard to measure.
In addition, the mass of the electron neutrino, which consists mainly
of eigenstates of mass $3\times 10^{-5}$ eV, is undetectable directly
or by neutrinoless double beta decay.  The latter process measures an
effective neutrino mass, but even contributions to that from the $\nu_\mu$
and $\nu_\tau$ must be so small as to make detection very unlikely, although
other conjectured processes unrelated to neutrino mass could cause this
decay.

On the other hand, the dimension size of 0.06 mm, suggested by the average
rates of the three types of solar experiments
should be detectable by gravity
experiments.  The present best limit\cite{Hoyle} on such effects is
less than a factor of four from that value.

The large size of the extra dimension raises issues about cosmological and
supernova limits from the effects of high KK states of both sterile
neutrinos\cite{barbieri,lukas,george} and gravitons\cite{Hanhart}, despite
uncertainties in the understanding of the complex regimes of the early
universe and the supernova core.  For sterile neutrino limits, this
phenomenology is aided because there is a single KK tower based on an
exceedingly small mass, the VO $\Delta m^2$ is an order of magnitude smaller
than usual, and for MSW the equivalent sin$^22\theta$ is more than an order
of magnitude smaller than for standard fits.  For the global B-L model, the
universe re-heat temperature could be very low ($> 0.7$ MeV works
cosmologically), reducing production of high KK states.  The high
string scale of the local B-L model would appear to avoid all these
constraints\cite{piai}, however.  More complete investigation of these
constraints may enable choosing between these quite different models,
both of which provide this new way to explain solar, atmospheric, and
LSND oscillation data and may give the first evidence for an extra large
dimension.

The work of R.N.M. is supported by a grant from the National Science
Foundation No. PHY-9802551.  The work of D.O.C. and S.J.Y. is supported
by a grant from the Department Of Energy No. DE-FG03-91ER40618.
We thank V. Barger, G. Fuller, W. Haxton, A.
Perez-Lorenzana, and Y. Totsuka for discussions.

\begin{table}[htbp]
\caption{Predicted seasonal variations in $\nu_e$ fluxes, excluding
the $1/r^2$ variation.  The model assumed
$\mu_0=0.32\times 10^{-2}$ eV, $m_0=0.34\times 10^{-4}$ eV, and
$\delta_{ee}=0.78\times 10^{-7}$ eV.}
\vskip 2mm
\begin{center}
\begin{tabular}{|l|c|c|c|}
$\theta-\theta_0$& Chlorine & Gallium & Water\\ \hline
0 (January 2)    & 0.3787   &  0.5144 & 0.4635 \\
$\pm \pi/2$      & 0.3762   &  0.5121 & 0.4633 \\
$\pi$ (July 4)   & 0.3747   &  0.5082 & 0.4631 \\
\end{tabular}
\end{center}
\label{Seasonal}
\end{table}

\end{document}